# Impact of the experimental approach on the observed electronic energy loss for light keV ions in thin self-supporting films

Barbara Bruckner[a,b,*], Philipp M. Wolf[a], Peter Bauer[a,b] and Daniel Primetzhofer[a]

[a] Department of Physics and Astronomy, Uppsala University, Box 516, 751 20 Uppsala, Sweden

[b] Johannes-Kepler University Linz, IEP-AOP, Altenbergerstraße 69, 4040 Linz, Austria

*barbara.bruckner@physics.uu.se

**Abstract**
Energy spectra of backscattered and transmitted ions with primary energies of 50 keV and 100 keV interacting with self-supporting foils were recorded with a Time-of-Flight Medium-Energy Ion Scattering setup in a single experiment. Self-supporting Au and W foils without backing material were used. For He ions transmitted through Au the spectrum of detected particles shows two distinct components corresponding to different energy losses in the film, whereas for protons no such phenomenon was observed. To determine the origin of these different contributions, measurements for different angles of incidence and scattering angles have been evaluated. The results suggest that the two components in the spectrum of transmitted He ions could be attributed to impact parameter dependent energy loss, being more prominent for He ions than for protons. The main origin of the necessary impact parameter selection along the different ion trajectories is expected to be texture in the Au-foils.

# 1. Introduction

Energetic ions moving in matter lose energy due to interaction with both electrons and nuclei of the target material. Commonly, this deceleration process is referred to as electronic $S_e$ and nuclear stopping $S_n$, respectively; the total stopping power $S = S_e + S_n$ is a measure for the mean energy loss per unit path length. Under the assumption that the energy transfer is governed by binary collisions with either a nucleus or an electron, $S$ can be written as [1]

$$S = S_n + S_e = n \int T(b) d\sigma(T) \quad \text{(Eq. 1)}$$

with $n$ being the atomic density of the target material, $b$ the impact parameter, $T(b)$ the impact parameter dependent energy transfer in a single collision and $d\sigma$ the corresponding differential scattering cross section. For convenience, typically, the stopping cross section $\varepsilon$ is used which is normalized to the atomic density in the target $n$: $\varepsilon = 1/n \cdot dE/dx$.

Information on the electronic energy loss in a sample of known thickness (typically given in at/cm$^2$) and composition can be deduced in two complementary experimental approaches: by detecting either (i) transmitted or (ii) backscattered projectiles with typical scattering angles of $\vartheta_t < 1°$ and $\vartheta_b > 150°$, respectively. In transmission geometry, the detected particles typically have undergone only small-angle deflections corresponding to large impact parameter collisions. In contrast, for a particle to be detected in backscattering geometry, at least one large-angle scattering event, i.e. small impact parameter, is necessary. This significant difference in the impact parameters probed along the trajectory of the projectile can in principle yield differences in the energy loss [2–4].

At high primary energies, where multiple scattering contributions can be neglected, a simple assumption is that in transmission geometry the projectile travels along a virtually straight line through the target without significant deflection. Analogously, also in backscattering geometry the trajectories of the projectile can be described by straight lines on the incoming and exiting path with an additional large-angle (back-)scattering event. The elastic energy loss in the large-angle collision does not contribute to the stopping power, since it is handled separately via the kinematic factor. Within this assumption, for both geometries the projectile is, for the majority of its total trajectory length traversing the electronic system at comparably large distances to the atomic nuclei. At lower energies multiple scattering has to be considered, which leads to an increase in the path length. However, the influence of multiple scattering along the trajectories of the projectiles is generally similar in transmission and in backscattering, except that in addition to small angle scattering for the latter only the large-angle scattering angle is not as well-defined as for higher primary energies [5].

In literature, for numerous target systems different sets of data for the electronic stopping cross section significantly differ from each other [6,7]. A possible explanation of the observed spread could be the different experimental approaches resulting in different contributions from impact parameter dependent energy loss. However, also other systematic uncertainties due to differences in the investigated targets such as crystallinity [8–12], composition [13] or inhomogeneity in thickness [14] could provide an explanation for the deviations.

So far, in joint studies between (i) HMI Berlin and JKU Linz [15,16], and (ii) the Instituto Atomico de Bariloche and JKU Linz [17] for primary energies around and below the stopping maximum no influence of different impact parameter dependence in proton stopping was observed within the experimental uncertainties. In both studies, the experiments were performed in different laboratories, i.e. scattering chambers, and in different approaches, transmission and backscattering. For the latter, both self-supporting foils and thin supported films were studied. The most crucial part for both experimental approaches is the use of targets of well-characterized composition, due to the potential contribution of surface and bulk impurities on the evaluation of electronic energy loss [18]. Also a preferential crystalline orientation in polycrystalline targets (texture) can affect the deduced electronic energy loss in either transmission or backscattering geometry. For projectiles with a primary energy of 300 keV and $Z_1 \geq 2$ transmitted through a polycrystalline Cu foil different energy loss components were observed in the recorded energy spectra which were attributed to different energy losses in channeling and random orientation [11]. In contrast, investigations of the electronic energy loss of protons and He ions in polycrystalline Au for primary energies ≤ 20 keV did not yield different energy loss components [19], although a small difference between the energy loss in channeled and random orientation was observed [10].

In this contribution, we show energy spectra of transmitted and backscattered particles from self-supporting foils obtained in the same scattering chamber. In this approach the data in transmission and backscattering are recorded subsequently with the primary beam on the same position of the sample, thereby allowing to exclude systematic uncertainties arising from different targets, irradiation spots, primary energy or energy calibration of the detector system. Polycrystalline gold and tungsten foils are chosen to cover an easy to handle high purity target as well as a reactive transition metal, with otherwise similar scattering kinematics. Additionally, for W electronic energy loss data for energies below the stopping maximum are scarce for both protons and He ions [6] although it is a material of high technological interest, e.g. as a first-wall material for fusion reactors [20].

## 2. Method

Energy spectra for backscattered as well as transmitted projectiles are recorded with a Time-of-Flight Medium-Energy Ion Scattering (Tof-MEIS) setup at the Tandem Laboratory at the Uppsala University [21,22]. The 350 kV Danfysik ion implanter can provide elemental and molecular ion beams between energies of 20 keV and 350 keV. The scattering chamber features a position-sensitive detector with a large solid angle (0.13 sr) which can be rotated around the sample resulting in scattering angles $\vartheta$ ranging from 0° to 160°. Therefore, both backscattering and transmission experiments can be performed with the same experimental setup with the position of the detector being the only changed parameter.

Two different types of samples were used in this contribution, i.e. custom purchased Au foils [23] and in-house manufactured W foils. For the W foils a layer of NaI was thermally evaporated on a Si substrate and subsequently, the W films were deposited by magnetron sputtering on the alkali halide layer using a MED 010 mini thin film deposition setup from Balzers. The alkali halide layer is dissolved in water and the floating W foil can be attached to the mounting ring. The areal density and purity of the self-supporting foils were measured using Rutherford Backscattering Spectrometry with a 2 MeV He$^+$ beam and a scattering angle of 170° and ToF-Elastic Recoil Detection Analysis (ERDA) measurements with a $^{127}$I$^{8+}$ beam and an incident and exit angle of 67.5°. The thickness in terms of the areal density as obtained from RBS experiments was converted to Å via the bulk densities for Au and W. For the Au foils areal thicknesses of 288 x 10$^{15}$ at/cm$^2$ and 135 x 10$^{15}$ at/cm$^2$ were obtained. For the thicker target, ERDA measurements indicate ~ 2.2 % O and 1.8 % C in the foil and assuming a bulk density of Au a thickness of 509 Å is obtained. Additionally, ~ 5 Å C are visible on both of the surfaces. The different sets of W foils contain a total amount of bulk impurities between 15 % and 25 % - typically, a mixture of C, O, N and H. The W sample presented in this contribution has a thickness of 652 Å W$_{0.76}$O$_{0.15}$C$_{0.09}$ with O and C as impurities. Both self-supporting foils are mounted on an aperture with a diameter of ~ 5 mm and beveled edges. In comparison, the typical beam size in MEIS is ~ 1x1 mm$^2$. Note, self-supporting W is very brittle, so the foils had to brought to vacuum quickly since they would not withstand long-term storage in air.

## 3. Results and discussion

In Fig. 1 Time-of-Flight to energy converted ToF-MEIS spectra for 50 keV protons scattered from a 509 Å Au foil are shown for (a) transmission at $\vartheta = 0° \pm 0.5°$ and (b) backscattering at $\vartheta = 160° \pm 2°$ geometry. Both spectra are obtained for normal angle of incidence, i.e. $\alpha = 0°$. In backscattering geometry, electronic energy loss can be deduced from the width of the metal signal. In transmission geometry information on the electronic stopping can be obtained from the position of the peak of transmitted particles $E_f$ (either the most probable or the mean energy loss) via the total energy loss $\Delta E$ in the film: $\Delta E = (E_0 - E_f)$. For low primary energies only small deviations (≤ 1.5 %) between evaluations of the mean or most probable energy loss were found [24]. In the spectrum of backscattered projectiles, the width $\Delta E'$ of the Au peak contains information on the total energy loss of the projectile in the film with $\Delta E' = k \cdot \Delta E_{in} + \Delta E_{out}$, with the kinematic factor $k$. Note, here both the energy loss on the incoming $\Delta E_{in}$ and exiting trajectory $\Delta E_{out}$ contribute.

In the presented energy regime, the enhanced scattering cross sections leads to significant contributions from multiple scattering hampering a straightforward evaluation of the electronic SCS. To properly consider multiple scattering the experimental spectra are compared to Monte-Carlo (MC) simulations. The employed TRBS code (TRIM for backscattering) allows to disentangle electronic and nuclear stopping [25]. The simulations included in this manuscript are performed using the thickness of the sample as well as bulk and surface contaminations as obtained from RBS and ERDA, and the TFM potential, which was found to yield good agreement for the multiple scattering background [26,27]. Convolution of simulated energy spectra by a Gaussian accounts for the finite energy resolution of the primary beam. The electronic SCS for both backscattering and transmission geometry is evaluated by varying the electronic stopping of the foil, $S_{e,foil}$, in the TRBS code until a best agreement between experiment and simulation is obtained. The electronic energy loss for Au, $S_{e,Au}$, is for both geometries evaluated from the energy loss in the foil, $S_{e,foil}$, by applying corrections according to Bragg's rule [13]. Note, this procedure yields a mean stopping cross section averaged over all relevant impact parameters with different impact parameter distributions for backscattering and transmission geometry.

The best fit of (a) the position and (b) the width of the Au peak is depicted as red solid line in Fig. 1. Additionally, in both plots a variation of the electronic energy loss of ± 3 % in the foil, $S_{e,foil}$, is plotted as red dashed lines. The electronic stopping cross section employed in the self-supporting film to achieve a best fit, is found equivalent within 3 % for both geometries and is in excellent agreement with reported data from literature, which were also deduced in backscattering geometry [27]. The observation of a minor discrepancy, i.e. an excess energy loss in transmission geometry, can at least be partially explained by surface contaminations such as adsorbed $H_2O$ or similar, which are stable under the present low-intensity beam. A different energy loss due to different impact parameters probed in transmission and backscattering, would, based on the assumption of higher energy loss at closer interaction distances between ion and target nuclei favor an increased energy loss in backscattering.

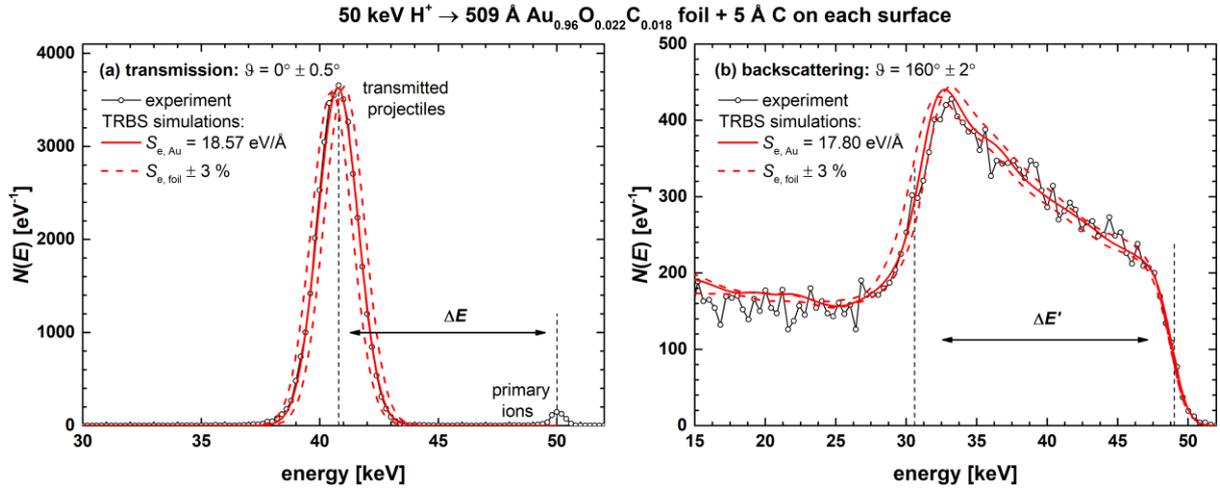

Figure 1: Energy spectra obtained in (a) transmission and (b) backscattering geometry for 50 keV $H^+$. The red solid lines represent TRBS simulations with a best fit to either spectrum, whereas the dashed lines correspond to a variation of the electronic energy loss of ± 3 %.

A comparison of energy spectra for $He^+$ ions is depicted in Fig. 2 obtained with a primary energy of 100 keV and the same scattering angles as in Fig. 1. Different from protons, the peak of transmitted projectiles in panel (a) features no single Gaussian distribution but shows a pronounced tail towards higher energies, i.e. projectiles which lost less energy than expected from an average trajectory in polycrystalline Au. The most probable energy loss is in good agreement with published data recorded in backscattering [27]. A similar contribution from trajectories with diminished energy loss is also observed in Au foils of different thicknesses and on different positions on the sample. In principle, this observation can be explained in different ways, e.g. by non-homogeneous foil thicknesses on lateral length scales significantly smaller than the beam spot. Alternative explanations could be impact parameter dependent energy loss as a consequence of partial channeling in structured Au foils, or fundamentally different trajectory types induced by other means, which contribute differently to the spectra in transmission and in backscattering, due to the significantly different impact parameter selection. Commonly, a gradual variation in the foil thickness that is visible in the transmission spectrum should also result in a broadened energy distribution for the trailing edge of the backscattering spectrum. However, in Fig. 2(b) no pronounced broadening of the trailing edge can be observed. The slight overestimation of the multiple and dual scattering background as well as the low-energy part of the spectrum could be only accounted for in a, rather unlikely model with two rather distinct thicknesses, which would reduce the signal from single scattering at larger depth and the corresponding background. To further exclude thickness inhomogeneities, for the energy spectrum obtained with 50 keV He$^+$ ions, simulations to fit both peak positions observed in the spectrum of transmitted He are performed with values for electronic stopping from literature [27] and the thickness of the sample as free parameter (spectrum not depicted here). The resulting thickness values are then used to simulate the spectrum of transmitted ions for H with a primary energy of 50 keV. Such simulations clearly show, that a simple thickness distribution necessary to explain the observation for helium would also yield two distinguishable peaks in the spectrum of transmitted protons, which is however, not observed. The two simulations plotted in Fig. 2 correspond to a best fit of the experimental data (for transmission the most probable energy loss) resulting in differences in the electronic energy loss within the experimental uncertainties, which can additionally be explained on a similar basis as for protons. Note, that also towards the trailing edge in the backscattering spectrum the simulation slightly overestimates the intensity of the spectrum, which could be missing intensity due to certain trajectories featuring an energy loss lower than average. However, also deficiencies of the scattering potential could in general explain a difference in the plateau height [26], while, the good agreement between simulation and experiment in the multiple scattering background renders this interpretation rather unlikely.

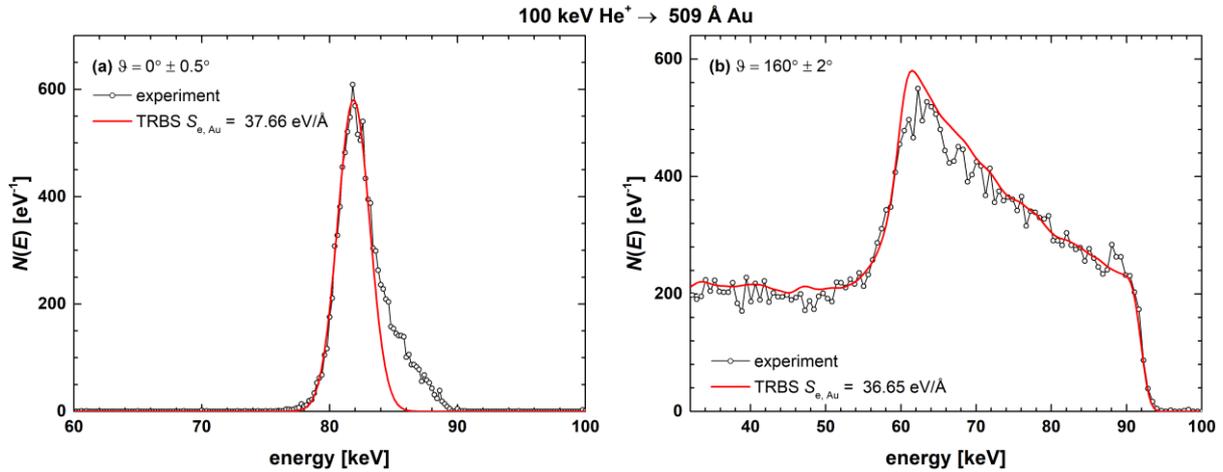

Figure 2: Energy spectra for 100 keV He$^+$ ions in (a) transmission and (b) backscattering geometry. The experimental details, as sample and scattering angles, are the same as in Fig. 1. Again, a comparison to TRBS simulations is shown as red solid lines.

To further investigate the high energy component in the spectrum of transmitted particles, different scattering geometries are studied. The large position-sensitive detector installed in the MEIS chamber allows for the evaluation of different regions of interest corresponding to scattering angles between 0° and ~ 12° obtained from the same incident beam in identical geometry. In Fig. 3(a) experimental spectra for one measurement obtained with 50 keV He$^+$ ions, $\alpha = 0°$ and different scattering angles are plotted. For all scattering angles $\vartheta \neq 0°$, an annulus around the primary ion beam has been evaluated for better statistics. Note, no difference in the spectrum of transmitted ions was observed depending on the azimuthal acceptance angle ($2\pi$ vs. 0.5°), as expected for polycrystalline or amorphous targets without azimuthal symmetry. With increasing scattering angles, the high energy contribution in the energy spectrum decreases in intensity and for $\vartheta \geq 5°$ it completely vanishes. In panel (b) and (c) different angles of incidence are chosen and evaluated for either $\vartheta = 0°$ or $\vartheta = 5°$, respectively. Again, an annulus corresponding to $\vartheta \pm 5°$ was evaluated, although due to the tilt of the sample the path lengths of the transmitted ions in the self-supporting foil are slightly different. Independent of the investigated angles of incidence the high energy contributions remains visible for scattering angles of 0°, however, its intensity diminishes with increasing $\alpha$. Similar to the spectra with normal incidence the high energy tail disappears for $\vartheta = 5°$. This decrease in the intensity of the high energy component as a function of the incidence angle is considered a strong indication for crystal texture in the Au foil combined with different specific energy loss in channeling and in random directions [10,11,28]. Even for the strongest inclination, the apparent thickness of the foil for traversing particles is increased by only ~ 7 %, which is not expected to strongly degrade the energy distribution of detected particles by either energy loss straggling or geometrical straggling. The observable shift in the energy spectra in Fig. 3 towards lower energies for increasing scattering and incident angles corresponds to larger energy losses due to the expected prolonged path length of the projectile. The observed increase in width in Fig. 3(c) can be associated with the different apparent film thicknesses resulting for given scattering angles as a consequence of the abandoned azimuthal symmetry. Note, that in the investigated energy regime, the possible difference in energy loss due to impact parameter dependent losses appears to be less pronounced for protons and therefore, with the given experimental resolution cannot be resolved in Fig. 1(a). In literature a more distinct difference in the energy loss between channeling and random geometry for He ions compared to protons observed for single crystalline target materials has been recently reported and attributed to energy losses due to charge exchange processes as well as charge state dependent stopping [8]. The proposed explanation fits also the observed inaccuracies in the fitting at energies around the trailing edge: different from a pronounced thickness inhomogeneity, such as a two-thickness model, partial channeling due to texture on the way to the large angle collision, will reduce the energy loss only on parts, i.e. at most half the trajectory, leading to smaller effects. In parallel, it will favor scattering at highest energies, i.e. close to the surface as well as slightly suppress plural and multiple scattering.

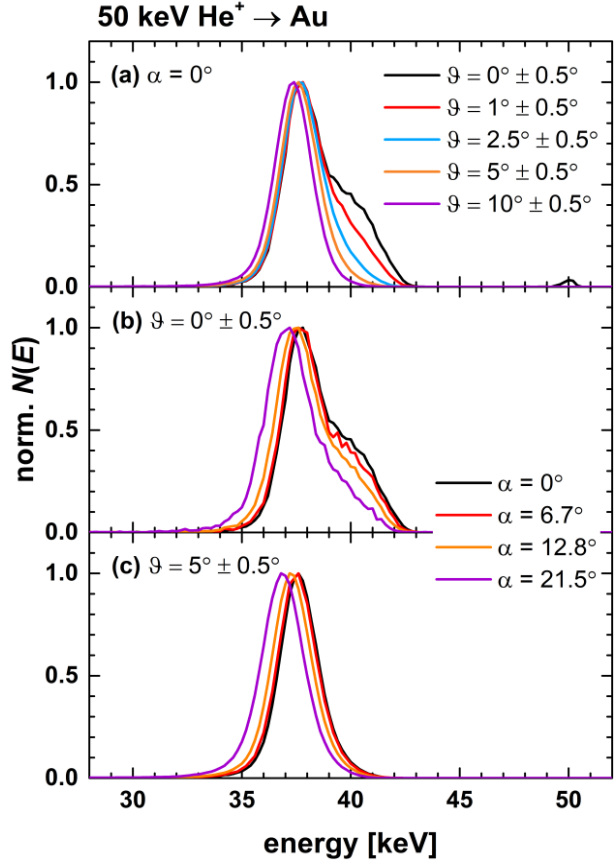

Figure 3: Experimentally obtained energy spectra for 50 keV He$^+$ ions transmitted through a Au foil with different angles of incidence $\alpha$ and scattering angles $\vartheta$: (a) normal incidence and different scattering angles evaluated at different regions of interest on the detector (one measurement), (b) $\vartheta = 0°$ and (c) $\vartheta = 5°$ with different angles of incidence. Note, the spectra in (b) and (c) for a certain incident angle are evaluated from one measurement.

Analogue to the Au foils transmission measurements are performed for a self-supporting W foil, which features significantly different properties compared to Au (structure, rigidity, impurities). In Fig. 4 energy spectra recorded for 50 keV He$^+$ ions transmitted through a W foil are presented, again for different scattering geometries and angle of incidence. Panel (a) shows different scattering angles $\vartheta$ for $\alpha = 0°$, whereas in panel (b) and (c) the incident angle was varied with $\vartheta = 0°$ and $\vartheta = 5°$, respectively. It can clearly be seen, that independent of the angle of incidence no contribution at high energy is found in the spectrum of transmitted particles for $\vartheta = 0°$. The absence of the high-energy component was also observed for an amorphous C foil with the same experimental parameters – 50 keV He$^+$ ions as projectile and $\alpha = 0°$ and $\vartheta = 0°$ (not depicted in this contribution). Consequently, as for self-supporting films of both carbon and a refractory metal such as tungsten with additional contaminations an amorphous or at least significantly less well-oriented crystal structure is expected, the observed effect for He in Au is likely to be caused by differences in energy loss depending on the probed impact parameters which are influenced by the crystallinity of the material featuring a textured surface. Similar observations have been reported for 300 keV He transmitted through a polycrystalline Cu foil [11].

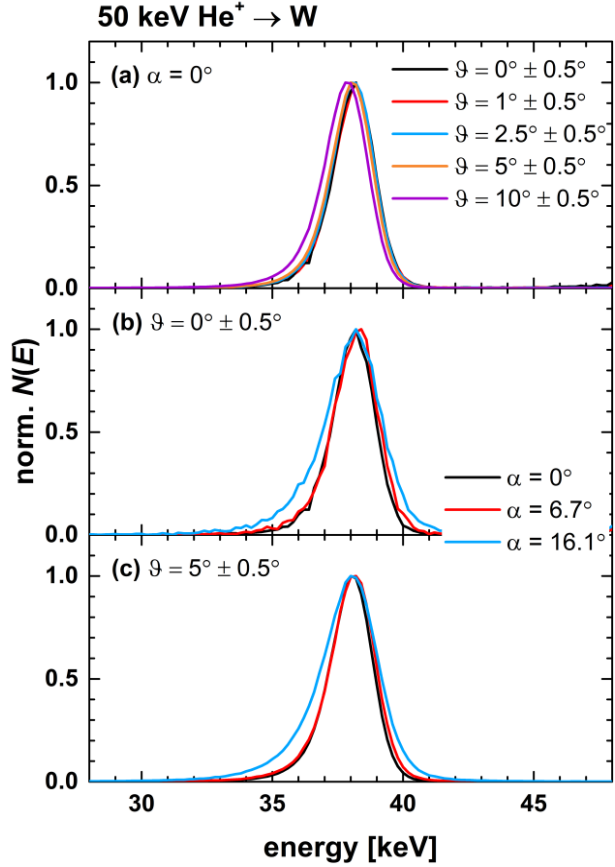

Figure 4: Energy spectra for 50 keV He$^+$ ions transmitted through a W foil, again for different incident $\alpha$ and scattering angles $\vartheta$ as in Fig. 3.

## 4. Summary and conclusion

In this contribution, we investigated energy spectra of protons and He$^+$ ions transmitted and backscattered from self-supporting Au and W foils. Both measurements were performed with similar experimental conditions, i.e. sample, ion beam, and detector. Only the position of the detector was varied between the measurements. For protons, the small deviations in the evaluated electronic energy loss between the different experimental geometries can mainly be attributed to surface contaminations on the self-supporting foils. Thus, for these projectiles and the investigated energies, no significant dependence of the deduced specific electronic energy loss on the specific experimental geometry is observable.

For He$^+$ ions transmitted through the Au foils two contributions in the energy spectrum could be observed, which was not the case for protons. By varying the scattering geometry (incident and exit angle) as well as the investigation of a different sample, the contribution at lower energy loss is attributed to effects of texture in the Au foil resulting in different components of the energy loss, as a result of trajectory selection and impact parameter dependent energy loss. Such a significant influence of the sample microstructure was also observed in a recent investigation of energy loss in single crystalline Si in both random and channeling geometry using both protons and He ions in a similar energy range [8].

Similarly as in the present study, the difference between the geometries was observed to be more pronounced for He ions compared to protons, which was attributed to charge state dependent stopping. Nevertheless, even for the presented complicated sample-microstructure, fitting the most probable energy loss in transmission and the width of the backscattering spectrum resulted in a specific electronic energy loss differing by less than 3 %. With slightly increased loss in transmission, this difference is very small compared to the observable spread of data from literature [6].

For typical non-channeling trajectories, as exemplified by experiments on less crystalline W-foils or by abandoning the condition of normal incidence, also for He ions, the observed specific energy loss in backscattering and transmission was found comparable within experimental uncertainties. The investigation of different combinations of target materials and projectiles can give more insight on details of the general relevance of the observed impact parameter dependence of electronic stopping for polycrystalline samples.

**Acknowledgment:**
Financial support by VR-RFI (contract #2017-00646_9) and the Swedish Foundation for Strategic Research (SSF, contract RIF14-0053) supporting accelerator operation is gratefully acknowledged. PB expresses his gratitude for the kind hospitality at UU.

**Bibliography**

[1]  P. Sigmund, Particle Penetration and Radiation Effects, Springer Verlag-Berlin Heidelberg, 2006.

[2]  R. Ishiwari, N. Shiomi, N. Sakamoto, Geometrical effect on the measurement of stopping power: Angle-dependent energy loss of 7-MeV protons in metallic and organic thin foils, Phys. Rev. A. 25 (1982) 2524–2528. doi:10.1103/PhysRevA.25.2524.

[3]  P. Sigmund, K.B. Winterbon, Small-angle multiple scattering of ions in the screened Coulomb region. I. Angular distributions, Nucl. Inst. Methods. 119 (1974) 541–557. doi:10.1016/0029-554X(74)90805-2.

[4]  P. Sigmund, A. Schinner, Note on measuring electronic stopping of slow ions, Nucl. Inst. Methods Phys. Res. B. 410 (2017) 78–87. doi:10.1016/j.nimb.2017.08.011.

[5]  D. Goebl, K. Khalal-Kouache, D. Roth, E. Steinbauer, P. Bauer, Energy loss of low-energy ions in transmission and backscattering experiments, Phys. Rev. A. 88 (2013) 032901. doi:10.1103/PhysRevA.88.032901.

[6]  H. Paul, C. Montanari, Electronic Stopping Power of Matter for Ions - IAEA Nuclear Data Services, Int. At. Energy Agency—Nuclear Data Serv. Vienna, Austria. (2016). https://www-nds.iaea.org/stopping (accessed February 9, 2018).

[7]  C.C. Montanari, P. Dimitriou, The IAEA stopping power database, following the trends in stopping power of ions in matter, Nucl. Inst. Methods Phys. Res. B. 408 (2017) 50–55. doi:10.1016/j.nimb.2017.03.138.

[8]  S. Lohmann, D. Primetzhofer, Disparate energy scaling of trajectory-dependent electronic excitations for slow protons and He ions, ArXiv. (2019). http://arxiv.org/abs/1907.08519 (accessed September 23, 2019).

[9]  J.E. Valdés, P. Vargas, C. Celedón, E. Sánchez, L. Guillemot, V.A. Esaulov, Electronic density corrugation and crystal azimuthal orientation effects on energy losses of hydrogen ions in grazing scattering on a Ag(110) surface, Phys. Rev. A. 78 (2008) 032902. doi:10.1103/PhysRevA.78.032902.

[10] R. Blume, W. Eckstein, H. Verbeek, K. Reichelt, Electronic energy loss of H, D and He in single crystal gold films in the energy range below 15 keV, Nucl. Inst. Methods. 194 (1982) 67–70. doi:10.1016/0029-554X(82)90491-8.

[11] P. Mertens, The influence of the polycrystalline structure of thin copper foils on the energy loss of transmitted 300 keV ions, Thin Solid Films. 60 (1979) 313–320. doi:10.1016/0040-6090(79)90077-4.

[12] E.S. Machlin, S. Petralia, A. Desalvo, R. Rosa, F. Zignani, Energy loss of protons channelling through very thin gold, Philos. Mag. 22 (1970) 101–116. doi:10.1080/14786437008228155.

[13] B. Bruckner, D. Roth, D. Goebl, P. Bauer, D. Primetzhofer, A note on extracting electronic stopping from energy spectra of backscattered slow ions applying Bragg's rule, Nucl. Inst. Methods Phys. Res. B. 423 (2018) 82–86. doi:10.1016/j.nimb.2018.02.005.

[14] I. Reid, P.J. Scanlon, Anomalous stopping power effects in thin gold films, Nucl. Inst. Methods. 170 (1980) 211–216. doi:10.1016/0029-554X(80)91014-9.

[15] P. Bauer, How to measure absolute stopping cross sections by backscattering and by transmission methods. Part I. backscattering, Nucl. Inst. Methods Phys. Res. B. 27 (1987) 301–314. doi:10.1016/0168-583X(87)90569-6.

[16] P. Mertens, How to measure absolute stopping cross sections by backscattering and transmission methods. Part II. transmission, Nucl. Inst. Methods Phys. Res. B. 27 (1987) 315–322.

doi:10.1016/0168-583X(87)90570-2.

[17] D. Roth, C.E. Celedon, D. Goebl, E.A. Sanchez, B. Bruckner, R. Steinberger, J. Guimpel, N.R. Arista, P. Bauer, Systematic analysis of different experimental approaches to measure electronic stopping of very slow hydrogen ions, Nucl. Inst. Methods Phys. Res. B. 437 (2018) 1–7. doi:10.1016/j.nimb.2018.09.028.

[18] P. Bauer, D. Semrad, P. Mertens, The influence of different experimental methods on the measured energy dependence of stopping powers, Nucl. Inst. Methods Phys. Res. B. 12 (1985) 56–61. doi:10.1016/0168-583X(85)90700-1.

[19] R. Blume, W. Eckstein, H. Verbeek, Electronic energy loss of H, D and He in Au below 20 keV, Nucl. Inst. Methods. 168 (1980) 57–62. doi:10.1016/0029-554X(80)91231-8.

[20] E. Bernard, R. Sakamoto, E. Hodille, A. Kreter, E. Autissier, M.F. Barthe, P. Desgardin, T. Schwarz-Selinger, V. Burwitz, S. Feuillastre, S. Garcia-Argote, G. Pieters, B. Rousseau, M. Ialovega, R. Bisson, F. Ghiorghiu, C. Corr, M. Thompson, R. Doerner, S. Markelj, H. Yamada, N. Yoshida, C. Grisolia, Tritium retention in W plasma-facing materials: Impact of the material structure and helium irradiation, Nucl. Mater. Energy. 19 (2019) 403–410. doi:10.1016/j.nme.2019.03.005.

[21] M.K. Linnarsson, A. Hallén, J. Åström, D. Primetzhofer, S. Legendre, G. Possnert, New beam line for time-of-flight medium energy ion scattering with large area position sensitive detector, Rev. Sci. Instrum. 83 (2012) 095107. doi:10.1063/1.4750195.

[22] M.A. Sortica, M.K. Linnarsson, D. Wessman, S. Lohmann, D. Primetzhofer, A versatile time-of-flight medium-energy ion scattering setup using multiple delay-line detectors, Nucl. Inst. Methods Phys. Res. B. 463 (2020) 16–20. doi:10.1016/J.NIMB.2019.11.019.

[23] Lebow Company, (n.d.). http://lebowcompany.com/ (accessed October 22, 2019).

[24] E.A. Figueroa, N.R. Arista, J.C. Eckardt, G.H. Lantschner, Determination of the difference between the mean and the most probable energy loss of low-energy proton beams traversing thin solid foils, Nucl. Inst. Methods Phys. Res. B. 256 (2007) 126–130. doi:10.1016/j.nimb.2006.11.103.

[25] J.P. Biersack, E. Steinbauer, P. Bauer, A particularly fast TRIM version for ion backscattering and high energy ion implantation, Nucl. Inst. Methods Phys. Res. B. 61 (1991) 77. doi:10.1016/0168-583X(91)95564-T.

[26] B. Bruckner, T. Strapko, M.A. Sortica, P. Bauer, D. Primetzhofer, On the influence of uncertainties in scattering potentials on quantitative analysis using keV ions, ArXiv. (2019) 1908.02045. http://arxiv.org/abs/1908.02045 (accessed September 4, 2019).

[27] D. Primetzhofer, Inelastic energy loss of medium energy H and He ions in Au and Pt: Deviations from velocity proportionality, Phys. Rev. B. 86 (2012) 094102. doi:10.1103/PhysRevB.86.094102.

[28] C.J. Andreen, R.L. Hines, Channeling of D+ and He+ ions in gold crystals, Phys. Rev. 151 (1966) 341–348. doi:10.1103/PhysRev.151.341.